\documentclass[apj]{emulateapj}
\usepackage{apjfonts}
\renewcommand\plotone[1]{\centering\includegraphics[width=\columnwidth]{#1}}
\submitted{Accepted for publication in ApJ Letters}

\shorttitle{A STRONG TEST OF ELECTRO-WEAK THEORY}
\shortauthors{WINGET ET AL.}

\begin{document}

\title{A Strong Test of Electro-Weak Theory using Pulsating 
DB White Dwarf Stars \\ as Plasmon Neutrino Detectors}

\author{D. E. Winget\altaffilmark{1},
        D. J. Sullivan\altaffilmark{2},
        T. S. Metcalfe\altaffilmark{3},
        S. D. Kawaler\altaffilmark{4},
        M. H. Montgomery\altaffilmark{5}}

\altaffiltext{1}{\footnotesize Department of Astronomy and McDonald 
                Observatory, The University of Texas, Austin, TX 78712}
\altaffiltext{2}{\footnotesize School of Chemical and Physical Sciences, 
                Victoria University of Wellington, Wellington, NZ}
\altaffiltext{3}{\footnotesize Harvard-Smithsonian Center for Astrophysics, 
                60 Garden Street, Cambridge, MA 02138}
\altaffiltext{4}{\footnotesize Department of Physics and Astronomy, 
                Iowa State University, Ames, IA 50011}
\altaffiltext{5}{\footnotesize Institute of Astronomy, University of Cambridge,
                Madingley Road, Cambridge CB3 0HA, UK}

\begin{abstract}
We demonstrate that plasmon neutrinos are the dominant form of energy loss
in model white dwarf stars down to $T_{\rm eff} \sim 25,000$~K, depending
on the stellar mass.  The lower end of this range overlaps the observed
temperatures for the V777 Her star (DBV) instability strip. The evolution
of white dwarfs at these temperatures is driven predominantly by cooling,
so this directly affects the stellar evolutionary timescale in proportion
to the ratio of the neutrino energy loss to the photon energy loss.  This
evolutionary timescale is observable through the time rate of change of
the pulsation periods. Although the unified electro-weak theory of lepton
interactions that is crucial for understanding neutrino production has
been well tested in the high energy regime, the approach presented here
should result in an interesting low-energy test of the theory. We discuss
observational strategies to achieve this goal.
\end{abstract}


\section{INTRODUCTION}

Not long after Fermi's theory of the weak interaction was first
generalized to include, among other things, the possibility of a direct
interaction between the electron and the relevant neutrino 
\citep[e.g.,][]{fg58}, a variety of theoretical papers pointed out that this
interaction, though extremely feeble, could have a major impact in the hot
dense plasmas to be found in the astrophysical domain 
\citep[see][ for an early review]{fh64}. However, given the weakness of
neutrino interactions, direct measurements of these rates in astrophysical
objects have proved a great challenge. These measurements rely on enormous
detecting volumes in order to obtain a significant number of events. In
fact, only in the case of the Sun and SN~1987A have direct links been made
between detections of neutrinos and particular astronomical objects.

It is possible to measure neutrino rates, albeit indirectly, in a
third class of objects, the white dwarf stars.  For both the hot white
dwarf and a pre-white dwarf stars the theoretical neutrino energy losses
exceed the photon energy losses, and so essentially control the
evolutionary timescale.  The astrophysical significance of neutrino
emission in hot pre-white dwarf stars has been investigated
extensively by many authors, building on the foundations of work
undertaken at the University of Rochester
\citep[e.g.][]{vil66,ks69,svv69}. The principal observable signature
of these neutrinos was thought to be in the white dwarf luminosity
function \citep[e.g., see][]{lv75}. In the mid-eighties we
\citep[e.g.,][]{kwh85} demonstrated that evolutionary frequency drifts
in the non-radial gravity modes present in pulsating hot pre-white
dwarf (DOV) stars might provide a very sensitive probe of plasmon
neutrino emissions in just the temperature range where the neutrino
luminosity was near its maximum. We later reported the observational
detection of an evolutionary rate of period change for one DOV,
PG~1159-035 \citep{win85,ckw99}, but the theoretical interpretation of
this result remains elusive \citep[see][ for a recent
summary]{ok00}. Theoretical models for PG~1159-035 have been unable to
reproduce the observed period structure and the rate of period change
simultaneously. Since the mechanical and thermal structure is more
closely coupled than in cooler white dwarf stars, it is more difficult
to identify the source of this problem.

No such difficulties should arise for the helium atmosphere DBV white
dwarfs, since their cooler temperatures lead to greater degeneracy and
hence a decoupling of their thermal and mechanical structure (see
section \ref{sec31} below). As a consequence, we should be able to
measure the plasmon neutrino flux from these objects, testing the
calculations of \citet{ito96} and measuring the net effect of plasmon
neutrinos.

\section{PLASMON NEUTRINOS}

Of the many possible neutrino emission processes, it is the plasmon
neutrino process that dominates neutrino production in hot white dwarf
interiors---the bremsstrahlung process comes a distant second. The
existence of plasmon neutrinos is an important test of our understanding
of the universality of the weak interaction. The nature of the process
that creates them is well described by \cite{cla68}, and we summarize it
here.

A free photon cannot decay into a neutrino and an anti-neutrino because
the constraint of coupling to a spin-1 particle requires the neutrinos to
be emitted in opposite directions; as a result, energy and momentum cannot
simultaneously be conserved and the process is forbidden. A photon
propagating in a plasma does not have this problem---it conserves both
energy and momentum by coupling to the plasma; such a coupled photon is
referred to as a ``plasmon''. A plasmon with sufficient energy can decay
into a neutrino and an anti-neutrino; the neutrinos created in this
process are termed plasmon neutrinos:
$$
\gamma^* \rightarrow \nu + \bar{\nu}.
$$
Under normal stellar conditions, including white dwarf interiors, we
expect the neutrinos created in this way to leave the star without further
interaction, resulting in net energy loss.

For a physical picture of what a plasmon actually is, we can consider the
photon to be a classical electromagnetic wave moving through a dielectric
medium.  If $\omega$ is the frequency of the photon, then we have
$$
\omega^2 = k^2 c^2 + \omega_0^2 ,
$$
where $\omega_0$ is the plasma frequency. Since $E=\hbar\,\omega$ and
$p=\hbar\, k$ for a photon, this can be re-written as
$$
E^2 = p^2 c^2 + m^2_{\rm pl} c^4 ,
$$
where
$$
m_{\rm pl} = \frac{\hbar \omega_0}{c^2}
$$
acts as the effective mass of the plasmon. Thus, the plasmon behaves like
a particle with a non-zero rest mass, and it is this effective mass that
enables it to decay into a neutrino and anti-neutrino pair.

The size of this ``rest mass'' energy, $\hbar\,\omega_0$, is important for
two reasons. First, in a general stellar environment these plasmon states
will be in thermal equilibrium with their surroundings, so significant
numbers of plasmons will be excited only if the typical thermal energy,
$k_{B}\,T$, is larger than the plasmon energy, $\hbar\,\omega_0$. Second,
the amount of energy liberated by the decay of a plasmon is related to its
mass, with high mass plasmons liberating more energy than low mass ones.
In low density non-degenerate gases, the plasma frequency is given by
$$
\omega_0^2 = \frac{4 \pi n_e e^2}{m_e} ,
$$
and we typically find that $\hbar\,\omega_0\ll k_{B}\,T$. Thus, even
though there are many plasmons, their rest mass energy is so small that
their decay releases little energy. In the dense plasmas of degenerate
stellar interiors the plasma frequency, now given approximately by
$$
\omega_0^2 = \frac{4 \pi n_e e^2}{m_e}
             \left[ 1 + \left(\frac{\hbar}{m_e c}\right)^2
             \left(3 \pi^2 n_e\right)^{\frac{2}{3}}\right]^{-\frac{1}{2}} ,
$$
is much larger, and plasmon neutrino production can lead to significant
energy loss. However, if the density is too large then only the highest
energy fluctuations out on the thermal tail will be large enough to excite
plasmons, leading to a suppression of energy losses through neutrino
emission.

\begin{figure}
\plotone{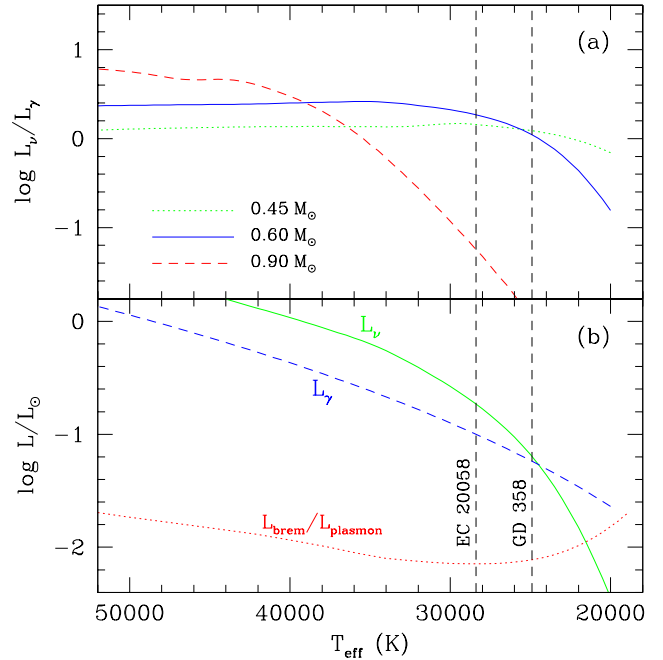}
\caption{(a) the ratio of neutrino to photon luminosity for white dwarf
models of various stellar mass.  In the region of interest (near the
temperatures of the pulsators GD~358 and EC~20058), neutrino losses for
the high-mass models are negligible, while they are very significant for
the less massive models. (b) The photon (dashed line) and neutrino (solid
line) luminosities for a $0.6\,M_{\odot}$ white dwarf, along with the
fractional contribution of bremsstrahlung neutrinos (dotted line).  The
neutrino luminosity is nearly twice the photon luminosity at the
temperature of EC~20058, and the two are approximately equal at the
temperature of GD~358.  The neutrino luminosity in this range is almost
entirely due to the plasmon reaction.\label{fig1}}
\end{figure}

This has important observational consequences. Theoretical evolutionary
calculations show that for temperatures greater than $T_{\rm eff} \sim
35,000\,$K (see Figure \ref{fig1}a), the ratio of neutrino to photon
luminosity is highest for massive white dwarf models. Higher plasma
frequencies result from higher central densities in more massive pre-white
dwarf models, yielding higher plasmon energies and correspondingly higher
neutrino energy fluxes. Eventually, as the models cool, the thermal
energies become too small for photons to excite plasmons.  This happens
first for the more massive model, and accounts for the fact that the
neutrino rates of the less massive models dominate for temperatures less
than 35,000~K. For the temperature range in which we are most interested,
the DBV instability strip, neutrino losses still dominate for models with
masses near the typical mass of the white dwarf stars, while they are
completely negligible for the more massive models.

\section{EVOLUTIONARY CALCULATIONS}

For the present work, we have computed several white dwarf evolutionary
sequences in order to explore the effects of plasmon neutrinos in the DBV
temperature range.  We have calculated sequences using polytropic starting
models as well as full evolutionary sequences calculated from various
main-sequence progenitor models, including different parametric treatments
of mass loss.  As has been well established by numerous authors
\citep[e.g.][]{kaw86}, the thermal history is erased by the time these
models cool to 35,000 K and the model sequences become indistinguishable
for our purposes. We have calculated sequences with and without plasmon
neutrino emission included, using the rates of \cite{ito96} in our models
(see Figure \ref{fig1}). At the energies relevant to our calculations, the
difference between these newer rates and the older \cite{bps67} rates are
$\sim$10 percent. For more complete details about the computation of the
models, see \cite{mnw00}.

\subsection{Period Structure \label{sec31}}

The problem of the coupling of the mechanical and thermal structure,
prominent in the DOVs, is nearly nonexistent by the time the models reach
the temperature domain of the DBV stars. For example, we calculated the
periods of models with $T_{\rm eff}\sim$ 28,000~K from two sequences: a
sequence including plasmon neutrino energy losses, and a sequence with the
neutrino production rates set to zero. The differences between the
corresponding pulsation periods of these models were always $\la$1 second,
with a typical difference of 0.24 seconds. We note that current
uncertainties in the interior mechanical structure, such as the detailed
C/O abundance profiles and the ``double-layered'' envelopes discussed by
\cite{fb02} and \cite{mmk03}, only affect the periods at this same level
of 1 second or less.

The energy loss to plasmon neutrinos primarily results in a reduced
central core temperature compared to models without neutrinos (for the
case considered above, this reduction in core temperature was $\sim$20
percent). The pulsation periods are only slightly affected because the
Brunt-V\"ails\"al\"a frequency is small in the core, where the pressure
support is dominated by the strongly degenerate electrons. As a
consequence, the g-mode pulsations are largely insensitive to the thermal
structure in the core, which is the only thing that changes dramatically
as a result of plasmon neutrino production.

\begin{figure}
\plotone{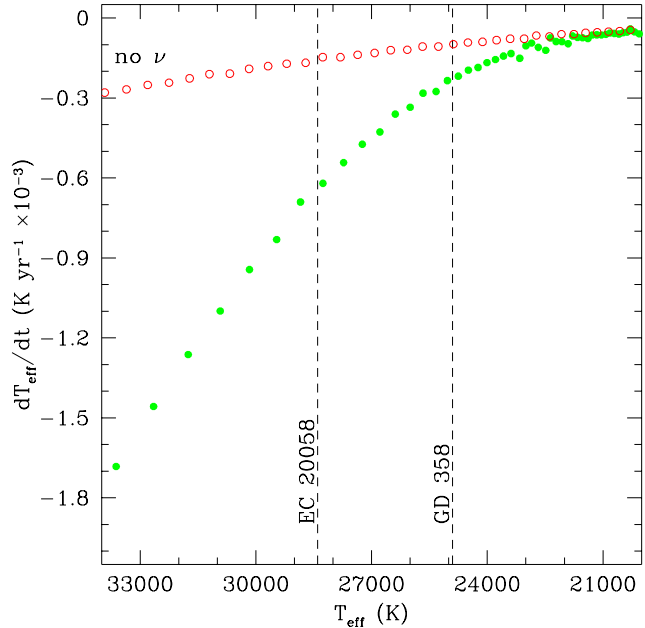}
\caption{The rate of change of the effective temperature for white dwarf
models that include (solid points) and ignore (open points)  neutrino
energy losses. The effective temperatures of two pulsating white dwarfs
are shown as dashed lines. The models differ by a factor of 4 at the
temperature of EC~20058 and a factor of 2 at the temperature of
GD~358.\label{fig2}}
\end{figure}

\subsection{Rates of Period Change}

Although the change to the thermal structure due to plasmon neutrinos does
not significantly alter the period structure, it does have a significant
effect on DBV models; it shortens the evolutionary timescale in the
temperature range 30,000~K $\ga T_{\rm eff} \ga$ 25,000~K. This is easily
seen in Figure \ref{fig2}, where we show the effect of plasmon neutrinos
on the time rate of change of the effective temperature. The dashed lines
indicate the temperature estimates of \cite{bea99} for two DBV stars: the
hotter one is EC~20058$-$5234, and the cooler one is GD~358. This should
correspond roughly to the hot half of the DBV pulsation instability strip
\citep{bw94a}.

Although the temperature of EC~20058 deduced by \cite{bea99} is relatively
uncertain due to a poor quality spectrum, a comparison of the observed
periods in the two stars provides independent asteroseismological evidence
that EC~20058 is significantly hotter than GD~358.  This is because the
dominant pulsation periods in a white dwarf should be tracers of the
thermal timescale at the base of the partial ionization/convection zone.
Since the observed periods are much shorter in EC~20058 than in GD~358,
EC~20058 should have a shallower convection zone, and hence a higher
effective temperature \citep[e.g., see][ and references therein]{win98}.

At these effective temperatures the evolution is dominated by cooling,
with gravitational contraction playing a diminishing role. The cooling
depletes the thermal energy stored in the ions. In models that include
neutrinos, the thermal energy deep in the core is further depleted by the
emission of plasmon neutrinos, and the evolutionary timescale is shorter
than in models that ignore the effects of neutrinos. Figure \ref{fig2}
makes it clear that the effect of the plasmon neutrinos on the evolution
is {\it not subtle}. Near the high temperature end of the observed DBV
instability strip the evolutionary timescales of the models differ by a
factor of about four, and by the middle of the instability strip they
still differ by a factor of two.

\begin{figure}
\plotone{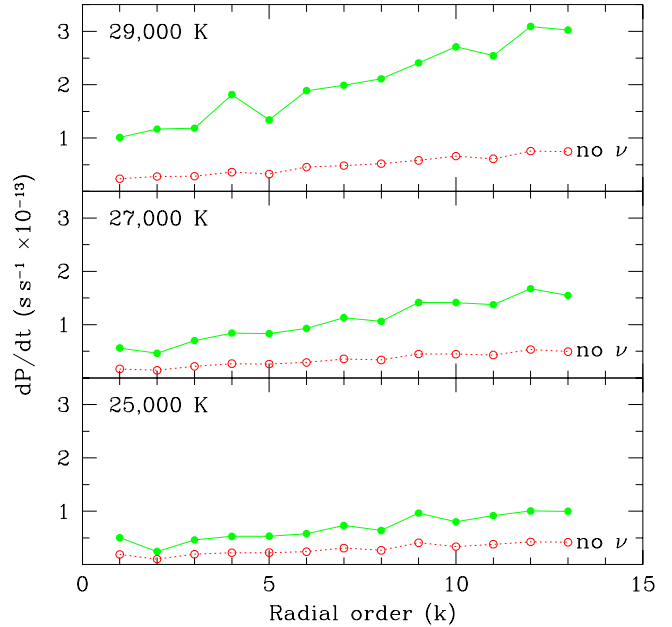}
\caption{The rate of change for pulsation periods in models on the hot end
of the DBV instability strip, including (solid points) and ignoring (open
points) the effects of neutrinos. An observational measurement of the rate
of period change for hotter DBV stars should allow us to detect the
signature of neutrino emission.\label{fig3}}
\end{figure}

This suggests that measurement of the evolutionary period change in hot
DBV white dwarf stars would make an excellent probe of the plasmon
neutrino production rates. As a proof of principle, we show in Figure
\ref{fig3} the rates of period change for theoretical evolutionary models
with temperatures in the hot half of the DBV instability strip.

We have already established that model uncertainties at the current level
do not alter the individual periods in a significant way for these
measurements, and Figure \ref{fig3} shows that exact radial order
identification is also unnecessary. If we know the observed period and the
spherical harmonic degree of a mode, then we can determine the radial
order within $\pm$1, which corresponds to a period range of 60--80~s for
models with masses near $0.6~M_\sun$. For a typical range of periods, this
gives us sufficient information to measure the plasmon neutrino production
rates with a precision of $\sim$10 percent.  We note that the these values
are only sensitive to spherical harmonic degree ($\ell$) at $\la 10$
percent: the difference between $\ell = 1$ and $\ell = 2$ for the same
period is $\sim$10 percent, and the difference between $\ell = 2$ and
$\ell = 3$ is $\la 1$ percent.

\section{DISCUSSION}

Our work demonstrates the potential of directly measuring the effect
of plasmon neutrino energy loss rates in the hot DBV white dwarf
stars.  Currently only one known DBV, EC~20058$-$5234, has both the
high temperature and period stability necessary to measure an
evolutionary period change.  After its discovery by \cite{koe95}, this
object was observed in a 1997 multi-site Whole Earth Telescope
campaign \citep[{\sc xcov}15,][]{sul03b}, with regular single-site
observations since then \citep{sul03a}. Work on this particular object
is in progress, but to fully exploit the potential of the DBV stars as
plasmon neutrino detectors we need to undertake a comprehensive
research program that includes the following components.

First, it will be important to expand the sample of known DBV stars well
beyond the 9 currently known.  This should provide access to a selection
of hot DBV pulsators that have the required period stability for measuring
evolutionary period changes. Fortunately, this goal appears to be feasible
using candidate white dwarfs identified in the Sloan Digital Sky Survey,
as has been adequately demonstrated by the recent large increase in the
number of known DAV stars using this source \citep{muk03}.

Second, we must obtain better effective temperatures for the hot DBV stars
of interest. Although not as important, it will also be useful to identify
the $\ell$ values of relevant pulsation modes. Appropriate HST
observations will be an important factor here. Both Figure \ref{fig2} and
Figure \ref{fig3} indicate that the more accurately we can determine the
temperature, either asteroseismologically, spectroscopically, or both, the
better limits we can place on dP/dt. The model rates of period change also
depend on the pulsation mode $\ell$ value, so knowledge of this index will
minimize the uncertainties.

Third, the overall program will be strengthened by access to a range of
DBVs with different masses. Figure \ref{fig1}a shows that massive white
dwarf stars ($\sim$0.9~M$_{\sun}$) are not producing significant plasmon
neutrinos in this temperature range, and so can serve as a control.  The
rates of period change for these stars should be insensitive to plasmon
neutrino production rates. In this way they will allow us to calibrate out
any possible systematic effects.  Comparative plasmon neutrino fluxes will
also allow us to calibrate the fluxes as a function of density, and
therefore plasmon energy.

Fourth, we need to undertake long-term photometric campaigns (in addition
to the on-going EC~20058 observations) on suitable DBV candidate stars. We
estimate that an observational timebase of between three and six years
duration, depending on individual timing accuracies, will be required to
achieve the goal of observing the predicted rates of period change.
Two-meter class telescopes and efficient frame transfer time-series
photometers will probably be required to undertake these campaigns, since
the newly detected DBVs will undoubtedly be, on average, significantly
fainter than those in the existing sample.

In summary, our work suggests that not only can we demonstrate the reality
of the plasmon neutrino process, but we can also quantitatively constrain
the production rates in the temperature-density domain relevant to white
dwarf interiors.

\acknowledgements
The authors thank S.J. Kleinman, A. Nitta, R.E. Nather, F. Mullally, and
A. Mukadam for many useful discussions. The work was supported in part by
grants NAG5-9321 from NASA's Applied Information Systems Research Program,
and ARP-0543 from the Texas Advanced Research Program.

\end{document}